\documentclass[12pt]{iopart}

\expandafter\let\csname equation*\endcsname=\relax
\expandafter\let\csname endequation*\endcsname=\relax

\usepackage{amsmath}

\usepackage{amssymb}

\usepackage{graphicx}

\usepackage{dcolumn}

\usepackage{bm}

\usepackage{xcolor}

\usepackage[export]{adjustbox}

\usepackage[T1]{fontenc}

\newcommand{\energy}{E}

\newcommand{\GkE}{{\cal{G}}^{\sigma}(\energy, \bm{k})}

\newcommand{\GkEup}  {{\cal{G}}^{\uparrow}  (\energy, \bm{k})}
\newcommand{\GkEdown}{{\cal{G}}^{\downarrow}(\energy, \bm{k})}

\newcommand{\GkEspinor}{{\cal{G}}(\energy, \bm{k})}

\newcommand{\sublattice}[1]{\tilde{#1}}

\newcommand{\eqdot}{\,\,\, .}

\newcommand{\eqcomma}{\,\,\, ,}

\newcommand{\MCAEnergy}{ {\cal{E}} }

\newcommand{\Operator}[1]
{
- \frac{1}{2 \pi N_{\bm{k}}} 
\; \times \\
&  \times
\sum_{ \sublattice{j} } \,
\int_{-\infty}^{\; \energy_{F}}
\mathrm{Im} \,
\mathrm{Tr_{L, \sigma}} \,
\Bigg\{
#1
\Bigg\}
\, dE 
}

\newcommand{\OperatorNoSublatticeNoTrace}[1]
{
- \frac{1}{2 \pi N_{\bm{k}}} 
\; \times \\
&  \times
\int_{-\infty}^{\; \energy_{F}}
\mathrm{Im} \,
\Bigg\{
#1
\Bigg\}
\, dE 
}

\newcommand{\PhiTetta}{\varphi, \theta}

\newcommand{\regexp}{\mathrm{exp}}

\newcommand{\regcos}{\mathrm{cos}}
\newcommand{\regsin}{\mathrm{sin}}

\newcommand{\EqRef}[1]{(\ref{#1})}

\newcommand{\Integral}{ \int_{-\infty}^{\; \energy_{F}} }

\newcommand{\TraceL}[1]
{
  \mathrm{Tr_{L}} \,
  \Bigg\{
    #1
  \Bigg\}
}

\begin{document}
       


\title[]
{
How ferromagnetic plane drives magnetocrystalline anisotropy in antiferromagnetic CoO and FeO
}



\author{Ilya~V.~Kashin}

\address
{
Theoretical Physics and Applied Mathematics Department, Ural Federal University, Mira Str. 19, 620002 Ekaterinburg, Russia
}

\ead{i.v.kashin@urfu.ru~(corresponding author)}


\author{Alexander~S.~Iakovlev}

\address
{
National Research Nuclear University MEPhI, Kashirskoe Shosse 31, 115409 Moscow, Russia
}

\ead{yas006@campus.mephi.ru}


\author{Sergei~N.~Andreev}

\address
{
Theoretical Physics and Applied Mathematics Department, Ural Federal University, Mira Str. 19, 620002 Ekaterinburg, Russia
}

\ead{s.n.andreev@urfu.ru}


\vspace{10pt}
\begin{indented}
\item[]February 2025
\end{indented}



\begin{abstract}
In this study we present a theoretical investigation of the role that ferromagnetic plane (111) plays in the formation of magnetocrystalline anisotropy (MCA) effects in CoO and FeO monoxides. For this purpose, a first-principles calculations of the electronic structure is performed within the GGA$+U$ approach. Based on the low-energy model in the Wannier functions basis, the MCA energy angular profile and the isotropic exchange environment of the transition metal atom are estimated using $\bm{k}$-dependent Green's functions. 
We have revealed a clear regularity in the direction of the easy and hard axes in both systems as lying in the (111) plane or along [111]. While for CoO the easy / hard axis orientation is (111) / [111], for FeO it appears reversed and thus emphasises the fundamental importance of (111) as the geometrical driver of the magnetism in the crystals.
The identification of the contributions that individual sublattices make to the MCA energy allowed us to reveal the decisive role of the electron hopping mechanisms in easy axis orientation. Considering the MCA and exchange environment with orbital decomposition in CoO and FeO under directional pressure in the (111) plane and along [111] showed a direct interrelation between the ferro- and antiferromagnetic contributions to the exchange environment and the energetic stability of the easy axis.
\end{abstract}



\vspace{2pc}






\section{Introduction}

At present, scientific interest both in the applied and fundamental fields of magnetism in solids is shifting towards compounds, possessed by non-trivial spin structures. 
For instance, such collective magnetic excitations as spin spirals and skyrmions are discovered to serve as the basic units in brand new computer memory modules with increased performance and low power consumption 
\cite{ohara2021confinement, ohara2022reversible, hassan2024dipolar}.

It is well known that the main generator of these structures is always an anisotropic mechanism of intra-atomic or inter-site nature. 
As the driving force for that one confidently considers the crystal symmetry, being reduced by external influence 
(e.g. pressure 
\cite{PhysRevMaterials.2.051004, D1DT01719E, craig2018probing}) 
or by a specific combination of internal factors.
Nevertheless, in order to make the lattice anisotropy influence the spin subsystem, a key condition is the presence of a significant spin-orbit coupling (SOC) on the magnetoactive atoms 
\cite{meng2016understanding, yuan2013spin, PhysRevB.79.224418, blonski2011magneto, bar2016magnetic}
or ligands 
\cite{C4CC05724D, parker2020ligand, goswami2012ligand, schweinfurth2017tuning, lado2017origin}.

One of the most common and anticipated effect of the type is magnetocrystalline anisotropy (MCA). It could be quite rigorously interpreted as the energy of the spin magnetic moment, which appeared dependent from the spatial orientation. Consequently, one can then speak about easy or hard axis (less often - plane), and the robustness of the effect could be defined as the corresponding energy difference.

Speaking about bulk crystals, we should note that expected values for this difference is about 1~$\mu$eV, which makes the MCA non-significant for the most particular cases. The basic reason one can state as relatively high crystal symmetry, if compared with quasi-two-dimensional materials where reduced crystal field naturally favors the presence of MCA as well as the magnetic shape anisotropy 
\cite{TwoDimensionalSystems_Qu, PhysRevMaterials.6.L061002}.
In this sense, the strong MCA $\sim$~1~meV in bulk compounds could be first predicted if layered structure takes place 
\cite{PhysRevApplied.3.034009, almasi2015enhanced, stohr1999exploring, shepley2015modification}.

\begin{figure}[htbp]
\begin{center}
\includegraphics[width=0.3\linewidth]{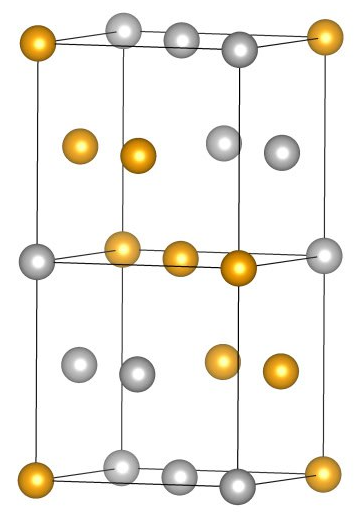} 
\caption{
Crystal structure of CoO and FeO.
Orange and gray spheres represent TM atoms, constituting ferromagnetic (111) planes with alternating direction of the local magnetic moments (AFM II).
}
\label{fig:CrystalStructure}
\end{center}
\end{figure}

And yet we can find representatives with no explicit accentuation of layers by means of the lattice. 
A striking example is iron and cobalt monoxides. 
Below the Néel temperature 
(293~K for CoO~\cite{bilz1979metal} and 
 198~K for FeO~\cite{10.1063/5.0082729}) 
the long-range antiferromagnetic order of AFM II type is stabilized in both compounds  
\cite{PhysRevB.86.115134, LIU201796, parida2018universality, deng2010origin}.
AFM II is assumed to be the stacking along cubic [111] axis of ferromagnetic planes with alternating direction of the Co(Fe) local magnetic moments 
(figure~\ref{fig:CrystalStructure}).
One considers the microscopic origin of this order as the superexchange interaction mediated by ligands 
\cite{PhysRevB.86.115134, PhysRev.79.350, ANDERSON196399, PhysRevB.80.014408}.

In context of MCA it appears essential because namely phase transition from the paramagnetic state to AFM II causes rhombohedral distortion. This distortion entails additional tetragonal 
\cite{PhysRev.110.1333, herrmann1978equivalent, PhysRevB.64.052102} 
or orthorhomic 
\cite{fjellvag2002monoclinic} 
distortions provided by Jahn-Teller effect, which has as the origin the degeneracy lifting between empty and filled $t_{2g}$ orbitals 
\cite{Jahn_Teller, Jahn_Bragg}.
It finally results in further reduction of the crystal symmetry to the monoclinic distortion and thus stimulates the strong MCA to emerge.

However, the microscopic origins of the easy axis formation are far from fully understood. 
Thus, despite the widespread popularity and extensive experience in the study of CoO and FeO, various theoretical and experimental works report different directions of this axis 
\cite{PhysRev.110.1333, herrmann1978equivalent, PhysRevB.64.052102, schron2013magnetic}.
It could be assumed to be another manifestation of the problem concerning known theoretical difficulties in accurate estimation of isotropic exchange interactions between nearest-neighbor transition metal (TM) atoms 
\cite{PhysRevB.80.014408, gopal2017improved}.
It results in that different methods of first-principles calculations produce significantly diverse magnetic pictures, 
with just reasonable agreement with the experimental data 
\cite{tomiyasu2006magnetic, PhysRevB.18.1317, rechtin1972long}.

The present work is then devoted to the identification of microscopic mechanisms generating MCA picture in CoO and FeO.
For this purpose we perform first-principles calculations within the framework of the GGA$+U$ method 
\cite{PhysRevLett.77.3865}, 
followed by the construction of a low-energy model of magnetoactive electron shell.
In order to estimate the energy of magnetic moment, localized on TM atom, with arbitrary spatial orientation Green's functions formalism was employed. Its advantage is known to address the crystal and electron structure of the values. 
In 
\cite{PhysRevB.52.13419, PhysRevB.89.214422} 
it is suggested on the base of inter-site Green's functions, and the MCA energy is defined as the direct spatial combination of pairwise contributions from each atom couple in the crystal. However, this combination appears practically able to be computed with reasonable accuracy only for insulating systems. 
If our aim is to study compounds with electron states near the Fermi level 
(CoO and FeO appear to be the case due to significant underestimation of the band gap reasoned by the strong localization of TM(3d) electrons and hybridization between TM(3d) and O(2p) states 
\cite{PhysRevB.86.115134, cinthia2015first}) 
then much more robust numerical stability is expected from using Green's functions with explicit dependence from the reciprocal space vector  
\cite{kashin2024magnetocrystalline}.

In 
\cite{kashin2024magnetocrystalline}  
by comparative analysis of the contributions, that individual transition metal sublattices in CoO and FeO make in forming of MCA energy, it was found that the dominant role is played by the ferromagnetic plane, containing the TM atom for which MCA energy is estimated.
In the present work we perform a deeper structural study of the MCA by implementing a structural decomposition of MCA energy value onto the orbital elements.
We confirm that inter-site and intra-atomic effects of TM atoms in ferromagnetic (111) plane plays decisive role in MCA picture of both materials. 
It appears particularly remarkable that while CoO(FeO) direction of the easy axis is reported as lying in plane(out-of-plane) 
\cite{kashin2024magnetocrystalline}, 
the hard axes in both compounds are found in the same geometry. 
We can then treat the orientation as in plane and out-of-plane with respect to (111).

Noteworthy that (111) is constituted by two sublattices: SL$_1$ (with the TM atom MCA energy is estimated for) and SL$_2$ (without this atom).
By comparing contributions of these sublattices in both compounds we reveal clear escalation of the role of SL$_2$ as one moves from hard to easy axis. 
It highlights the inter-site physics to be actually essential in MCA picture, on the background of SL$_1$, which captures both inter-site and intra-atomic effects.
Thus we provide a quantitative evidence towards the kinetic origin of MCA, having as a precursor superexchange interaction, which stabilizes AFM II magnetic order and consequently reduces the crystal symmetry.

In order to clarify the role of this interaction we consider CoO and FeO under directional pressure in (111) and along [111]. 
In the first case MCA is significantly suppressed, whereas in the second case it is even enhanced. Thus we provide new reliable evidence towards dominant role of the ferromagnetic (111) plane as the lattice feature, which is observed sensible to the external mechanical influence.
The orbital decomposition of the MCA energy and isotropic exchange environment, estimated in terms of infinitesimal spin rotations technique 
\cite{LIECHTENSTEIN198765, PhysRevB.106.134434}, indicate the direct relation between the ferro- / antiferromagnetic tendencies in the exchange interactions and the robustness of the easy axis. It opens up new prospects for the magnetism in CoO and FeO as being experimentally manipulated by the crystal field, if directional pressure is applied or a thin film is designed.







\section{Method}

Our consideration starts with the low-energy model of magnetoactive electron shell. It was parametrized by projection of the wave functions, obtained from the first-principles calculations, into the basis of the Wannier functions. 
We express this model as its Hamiltonian, defined in the tight binding approximation
\begin{equation}
\label{Eq:TBModel}
      \hat{H} =
            \sum_{i \ne j}
                  \sum_{\alpha \beta}
                        \sum_{\sigma}
                          \, t^{\sigma}_{i(\alpha) \, j(\beta)}
                                    \,
                              \hat{c}^{\dagger}_{i(\alpha) \sigma}
                                        \hat{c}_{j(\beta)  \sigma}
             + 
            \sum_{i}
                  \sum_{\alpha}
                        \sum_{\sigma}
                           \, \varepsilon^{\sigma}_{i(\alpha)}
                                    \,
                              \hat{c}^{\dagger}_{i(\alpha) \sigma}
                                        \hat{c}_{i(\alpha) \sigma} \eqcomma
\end{equation}
where 
$i$, $j$ are atom indices;
$\alpha$, $\beta$ are orbital indices;
$\sigma = \uparrow, \downarrow$ is spin index;
$\hat{c}^{\dagger}_{i(\alpha) \sigma}$,
$\hat{c}_{j(\beta) \sigma}$ are creation and annihilation operators;
$\varepsilon^{\sigma}_{i(\alpha)}$ is the on-site electron energy;
$t^{\sigma}_{i(\alpha) \, j(\beta)}$ is the electron hopping integral.

If one presents this Hamiltonian in a matrix form
$H^{\sigma}  ( \bm{T} )$ 
($\bm{T}$ being a translation vector), 
then the size of this matrix is defined by  the number of atoms in the unit cell and amount of orbitals in these atoms.
As we move from real to reciprocal space, the Hamiltonian matrix of the same size $H^{\sigma}(\bm{k})$ 
($\bm{k}$ is the reciprocal space vector of the Monkhorst-Pack grid \cite{PhysRevB.13.5188})
thus captures the entire sublattices of the crystal structure instead of individual atoms.
The sublattice (SL) should be understood as conditionally infinite group of atoms from all considered unit cells (one from each cell), where all atoms have the same local positions in the unit cells. 
In case of CoO and FeO (4 TM atoms in the unit cell) there are 
two sublattices (SL$_1$ and SL$_2$), which constitute one ferromagnetic plane (111), and other two sublattices (SL$_3$ and SL$_4$), which compose the neighboring (111) planes with alternating directions of the local magnetic moments (figure~\ref{fig:CrystalStructure}). 
In our notations individual atom of a crystal is labeled as $i$, thus the sublattice that contains the atom is denoted as $\sublattice{i}$. 
Therefore, one can say that $H^{\sigma}(\bm{k})$ describes both intra-atomic effects and  interactions between sublattices in the form of corresponding sectors 
$\big[ H^{\sigma}(\bm{k}) \big]_{\; \sublattice{i} \sublattice{j}}$.

The same framework of sublattices is preserved at the level of Green's function, which could be basically defined as
\begin{equation}
\label{Eq:GreenKdependent}
      \GkE = \big\{ \energy - H^{\sigma}(\bm{k}) \big\} ^{-1} \eqcomma
\end{equation}
where 
$\energy$ is the diagonal matrix of the sweep energy 
with small imaginary part $i 0^{+}$.

\subsection
{
\label{Subsec:MCA}
Magnetocrystalline anisotropy energy
}

In order to introduce SOC on TM atoms we employ a second-order perturbation theory 
(the first-order term is equal to zero due to time reversal symmetry) 
\cite{PhysRevB.52.13419, PhysRevB.89.214422, PhysRevB.39.865, PhysRevB.47.14932, Goringe_1997} 
around the Hamiltonian \EqRef{Eq:TBModel}.
The obtained expression for MCA energy, that magnetic moment of the atom $i$ has if directed along
$ (\sin{\theta} \cos{\varphi}, \,
 \sin{\theta}   \sin{\varphi}, \,
 \cos{\theta} ) $,
can be written as \cite{kashin2024magnetocrystalline}
\begin{equation}
\begin{split}
\label{Eq:FinalMAE}
    \MCAEnergy_{i} \; & (\PhiTetta) =  
            \Operator
            {
                    \sum_{ \bm{k} } \,
                        H^{so}(\PhiTetta)  \,\,
                        \big[ \GkEspinor \big]_{\; \sublattice{i} \sublattice{j}} \,\,
                        H^{so}(\PhiTetta)  \,\, 
                        \big[ \GkEspinor \big]_{\sublattice{j} \sublattice{i}} \,
            }
        \eqdot
\end{split}
\end{equation}
where 
$\energy_{F}$ is the Fermi energy,
$\mathrm{Tr_{L \, \sigma}}$ is the trace over orbital ($\mathrm{L}$) and spin ($\mathrm{\sigma}$) indices, and
$N_{\bm{k}}$ is the number of Monkhorst-Pack grid points. 
We note that Green's functions here are written in a spinor form 
$\GkEspinor =
\Big(
\begin{smallmatrix}
    \GkEup   &          0             \\
          0             &    \GkEdown
\end{smallmatrix}
\Big)
$ .
In both compounds under our consideration are equal TM atoms, which makes the SOC operator for the TM $3d$ shell
$H^{so}~=~
\lambda \, 
\bm{{\cal{L}}} 
\bm{{\cal{S}}}~=~
\Big(
\begin{smallmatrix}
    [H^{so}]^{\uparrow \uparrow}     &    [H^{so}]^{\uparrow \downarrow}      \\
    [H^{so}]^{\downarrow \uparrow}   &    [H^{so}]^{\downarrow \downarrow}
\end{smallmatrix}
\Big)
$
equal for any atom 
($\lambda$ is the small parameter, $\bm{{\cal{L}}}$ and $\bm{{\cal{S}}}$ are the orbital momentum and the spin of the $3d$ shell).
The rotation from initial orientation $(0, 0, 1)$ 
to arbitrary 
$ (\sin{\theta}  \cos{\varphi}, \,
   \sin{\theta}  \sin{\varphi}, \,
   \cos{\theta} ) $ 
could be accomplished using 
$H^{so}(\PhiTetta) = {\cal{U}}^{-1}(\PhiTetta) \,\, H^{so} \,\, {\cal{U}}(\PhiTetta)$, where
\begin{equation}
    {\cal{U}}(\PhiTetta) = 
    \begin{pmatrix}
    \phantom{-} \regcos{(\theta / 2)} \phantom{ \cdot e^{i \varphi} } &
                \regsin{(\theta / 2)} \cdot e^{-i \varphi}              \\ 
  - \; \regsin{(\theta / 2)} \cdot e^{\; i \varphi} & 
        \regcos{(\theta / 2)} \phantom{ \cdot e^{-i \varphi} }
    \end{pmatrix}
\end{equation}
is the Wigner's rotation matrix.

Important to mention the option of \EqRef{Eq:FinalMAE} to readily distinguish contributions of different sublattices \cite{kashin2024magnetocrystalline}.
In the present work, we perform more detailed structural decomposition of the MCA energy value into spin and orbital terms.
For this purpose we use the specific property of matrix trace operation if applied to the product of two square matrices with size $N \times N$  
\cite{PhysRevB.106.134434, PhysRevB.91.125133}
\begin{equation}
    \mathrm{Tr}[X \cdot Y] = 
      \sum^{N}_{m = 1}
      \sum^{N}_{l = 1} 
          X_{ml} \; Y_{lm} 
                        = \sum^{N}_{m, \; l = 1} Z_{ml} \eqdot
\end{equation}

It results in the expression for the contribution to the MCA energy of TM atom $i$ from the sublattice $\sublattice{j}$ 
\begin{equation}
\begin{split}
\label{Eq:MCADecomposition}
   \{ & \MCAEnergy_{i, \; \sublattice{j}} \; (\PhiTetta) \}^{\alpha \beta} =  
            \OperatorNoSublatticeNoTrace
            {
                    \sum_{ \bm{k} } 
                        \Big(
                            H^{so}(\PhiTetta)  \,\,
                            \big[ \GkEspinor \big]_{\; \sublattice{i} \sublattice{j}} 
                        \Big)^{\alpha \beta}
                                   \cdot
                        \Big(
                            H^{so}(\PhiTetta)  \,\, 
                            \big[ \GkEspinor \big]_{\sublattice{j} \sublattice{i}} 
                        \Big)^{\; \beta \alpha}
            }
        \eqcomma
\end{split}
\end{equation}
where the indices $\alpha$ and $\beta$ denotes the spinor element. 
Hence sum of this components rigorously reproduces the total MCA energy value
\begin{equation}
    \MCAEnergy_{i} \; (\PhiTetta) = 
       \sum_{ \sublattice{j} } 
       \sum_{\alpha \beta} 
           \,
           \{ \MCAEnergy_{i, \; \sublattice{j}} \; (\PhiTetta) \}^{\alpha \beta}
 \eqdot
\end{equation}

\subsection{Isotropic exchange environment}

To provide an estimation of the isotropic exchange environment of any atom of the crystal the common practice is to employ Andersen's ''local force theorem'' 
\cite{LocalForceTheorem_1, LocalForceTheorem_2, Lichtenstein2013correl13} 
in terms of infinitesimal spin rotations approximation 
\cite{LIECHTENSTEIN198765}.
It provides an opportunity to estimate Heisenberg exchange interaction between arbitrary chosen couple of atoms as
\begin{equation}
\label{Eq:LKAG}
J_{ij} =
  \frac{1}{8 \pi} 
  \Integral
  \mathrm{Im} \,
  \TraceL
  {
    \sum_{ \sigma } \;
    \Delta_{i}  
    G^{\sigma}_{ij}
    \Delta_{j}  
    G^{- \sigma}_{ji}
  }
  \, d\energy
  \eqcomma
\end{equation}
where 
$\Delta_{i} =
 [H^{\uparrow}  ( \bm{T} = 0 )]_{ii} - 
 [H^{\downarrow}( \bm{T} = 0 )]_{ii}$ is the intra-atomic spin splitting 
and 
 $G^{\sigma}_{ij}$ is the inter-site Green's function:
\begin{equation}
\label{Eq:InterSiteGF}
  G^{\sigma}_{ij} = 
    \frac{1}{ N_{\bm{k}} }  \,
    \sum_{ \bm{k} }         \,
    [\GkE]_{\sublattice{i} \sublattice{j}}
      \cdot
    \regexp
       \big\{
      -i \bm{k} (\bm{T}_{j} - \bm{T}_{i})
       \big\}
     \eqdot
\end{equation}

Then the contribution of the sublattice $\sublattice{j}$ in the isotropic exchange environment of the atom $i$ could be formally accumulated as
\begin{equation}
\label{Eq:ExchangeEnvironment}
{\cal{J}}_{i, \; \sublattice{j}} = \sum_{j \ne i, \; \in \; \sublattice{j}} \; J_{ij}
\eqcomma
\end{equation}
where positive(negative) values indicate the tendency towards AFM II (FM) order in CoO and FeO.
However, in the presence of electronic states near the Fermi level the convergence of this sum is expected to be slow or even absent, with no reliable criterion 
\cite{PhysRevB.106.134434, PhysRevLett.116.217202}.
To solve this problem in 
\cite{PhysRevB.106.134434} it was proposed to consider \EqRef{Eq:LKAG} in the reciprocal space by performing the Fourier transform
\begin{equation}
\label{Eq:JqFourer}
[J(\bm{q})]_{\sublattice{i} \sublattice{j}} = 
\sum_{ j \; \in \; \sublattice{j} }
J_{ij} 
      \cdot
      \regexp
         \big\{
           i \bm{q} (\bm{T}_{j} - \bm{T}_{i})
         \big\}
\eqcomma
\end{equation}
where $\bm{q}$ is the reciprocal space vector and $\bm{T}_{i} = 0$ due to consideration of the $i$ atom always in the ''central'' unit cell.
We emphasize that $[J(\bm{q})]_{\sublattice{i} \sublattice{j}}$ could be estimated directly using $\bm{k}$-dependent Green's functions \EqRef{Eq:GreenKdependent} as 
\cite{PhysRevB.106.134434} 
\begin{equation}
\begin{split}
[ & J(\bm{q})]_{\sublattice{i} \sublattice{j}} =
  \frac{1}{8 \pi N_{\bm{k}}} 
  \times \\ &\times
  \Integral
  \mathrm{Im} \,
  \TraceL
  {
      \sum_{ \sigma } 
      \sum_{ \bm{k} }               \,
      \Delta_{i} \;
     [{\cal{G}}^{\sigma} (\energy,  \;  \bm{k + q})]_{\sublattice{i} \sublattice{j}}   \;
      \Delta_{j} \;
     [{\cal{G}}^{- \sigma} (\energy,  \bm{k})]_{\sublattice{j} \sublattice{i}}   \;
  }
  \, d\energy
  \eqdot
\end{split}
\end{equation}

Therefore, for the exchange environment \EqRef{Eq:ExchangeEnvironment} we should address only one vector $\bm{q} = 0$ and write
\begin{equation}
\label{Eq:ExchangeEnvironmentThroughJq}
{\cal{J}}_{i, \; \sublattice{j}} = 
   [J(\bm{q} = 0)]_{\sublattice{i} \sublattice{j}}
    - 
   J_{ii} \cdot \delta_{\sublattice{i} \sublattice{j}}
\eqcomma
\end{equation}
where intra-atomic parameter $J_{ii}$ could be found using \EqRef{Eq:LKAG}.
The orbital decomposition of \EqRef{Eq:ExchangeEnvironmentThroughJq} could performed similarly as it was showed in subsection~\Ref{Subsec:MCA} 
and presented in 
\cite{PhysRevB.106.134434}:
\begin{equation}
\label{Eq:JqDecomposed}
\begin{split}
\{ [ & J(\bm{q})]_{\sublattice{i} \sublattice{j}} \}^{\alpha \beta} =
  \frac{1}{8 \pi N_{\bm{k}}} 
  \times \\ &\times
  \Integral
  \mathrm{Im} \,
     \Bigg\{
      \sum_{ \sigma } 
      \sum_{ \bm{k} }               \,
 \Big(
      \Delta_{i} \;
     [{\cal{G}}^{\sigma} (\energy,  \;  \bm{k + q})]_{\sublattice{i} \sublattice{j}}
 \Big)^{\alpha \beta}
    \cdot
 \Big(
      \Delta_{j} \;
     [{\cal{G}}^{- \sigma} (\energy,  \bm{k})]_{\sublattice{j} \sublattice{i}} 
 \Big)^{\beta \alpha}
      \Bigg\}
  \, d\energy
  \eqcomma
\end{split}
\end{equation}

\begin{equation}
\label{Eq:JiiDecomposed}
\{ J_{ii} \}^{\alpha \beta} =
  \frac{1}{8 \pi} 
  \Integral
  \mathrm{Im} \,
      \Bigg\{
    \sum_{ \sigma } \;
 \Big(
    \Delta_{i}  
    G^{\sigma}_{ii}
 \Big)^{\alpha \beta}
     \cdot
 \Big(
    \Delta_{i}  
    G^{- \sigma}_{ii}
 \Big)^{\beta \alpha}
      \Bigg\}
  \, d\energy
  \eqdot
\end{equation}







\section{Results and Discussion}

Computational details of 
the first-principles calculations,
low-energy model construction 
and
MCA energy estimation 
are summarized in \ref{AppendixA}.
Obtained Wannier functions describe  
along with TM($3d$) states 
also 
TM($4s$), TM($4p$),
O($2s$) and O($2p$) states due to their strong entanglement.
Despite the possibility to reproduce the spin magnetic moments and lattice constants of CoO and FeO in a good agreement with the experimental measurements 
\cite{PhysRevB.30.4734, PhysRevB.49.10170, PhysRevB.74.155108}, DFT calculations are known to severely underestimate the band gap up to its complete disappearance. The reason lies in a field of significant hybridization of the TM($3d$) and O($2p$), complemented by strong localization of TM($3d$) electrons 
\cite{PhysRevB.86.115134, cinthia2015first}.
As expected, our results demonstrate the same tendency, leading to the electronic states near the Fermi level.
It does not appear as the methodological difficulty to study microscopic magnetic effects in these compounds 
\cite{PhysRevB.86.115134}.
However, it makes the inter-site Green's functions-based approaches for estimation the MCA energy and isotropic exchange interactions practically inapplicable due to the absence of the intrinsic convergence criterion for spatial sum over the atoms 
\cite{PhysRevB.52.13419, PhysRevB.89.214422, PhysRevB.106.134434, PhysRevLett.116.217202}. 
Thus the employment of $\bm{k}$-dependent Green's functions is established as the only way to perform the numerical estimations with a reliable accuracy.

\begin{figure}[htbp]
\begin{center}
\includegraphics[width=0.9\linewidth]{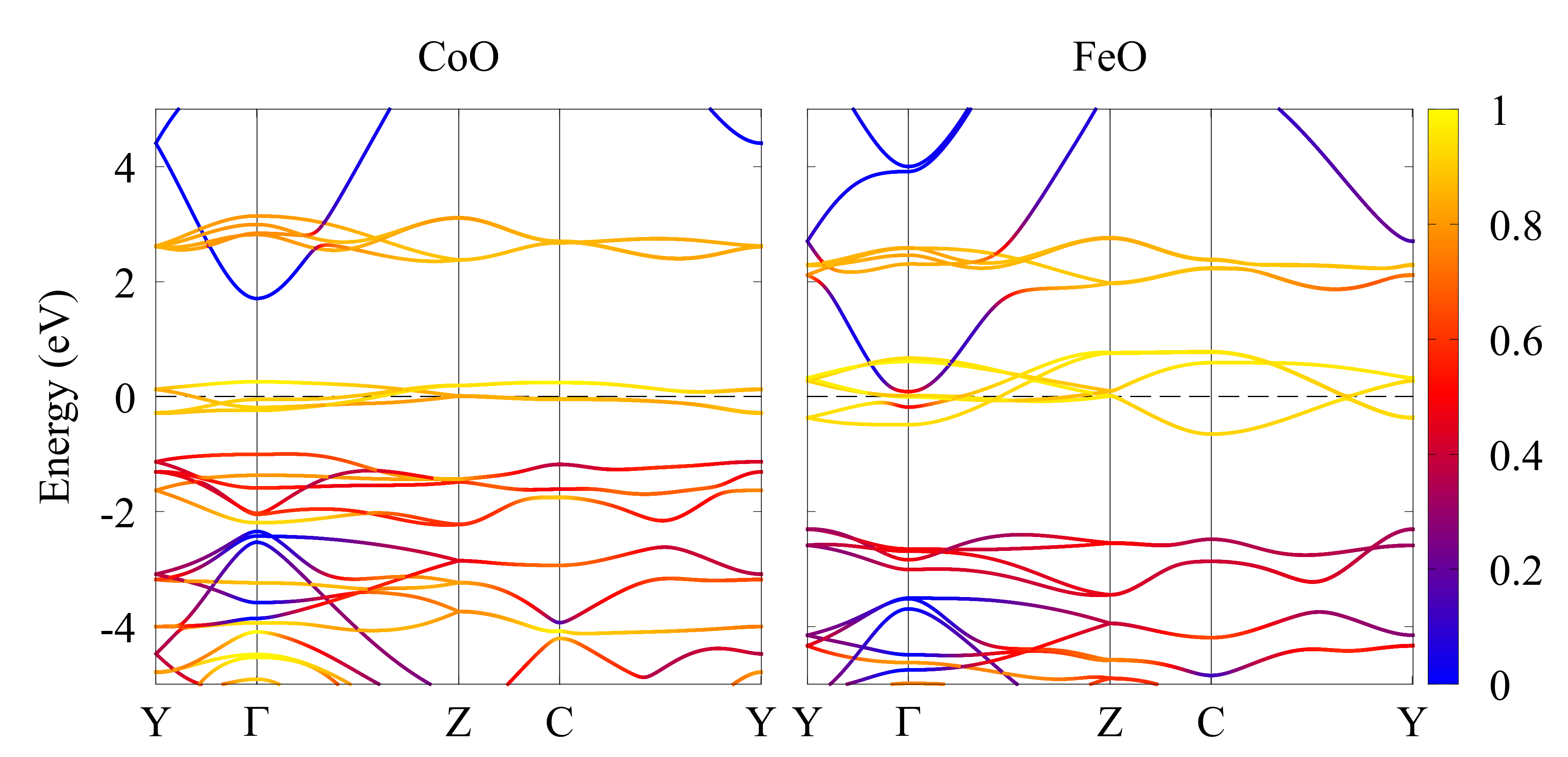} 
\caption{The band structure of CoO and FeO, obtained from GGA\textit{+U} calculations and further projection into a basis of Wannier functions. 
The high symmetry points are 
 $Y(0.5, 0, 0)$,
 $\Gamma(0, 0, 0)$,
 $Z(0, 0.5, 0)$, 
 $C(0.5, 0.5, 0)$.
The Fermi level is zero. 
The relative contribution of TM($3d$) states is depicted by color.}
\label{fig:BandStructure}
\end{center}
\end{figure}

In figure~\ref{fig:BandStructure} 
we present the band structure of the low-energy model, which perfectly reproduces initial DFT result in relevant energy range.
In order to distinguish the relative contribution of TM($3d$) states for each $\bm{k}$ vector  one should calculate the eigenvectors of $H^{\sigma}(\bm{k})$ 
and then find the ratio 
$F_{TM} / F_{All}$,
where 
$F_{TM}$ is the sum squares of moduli elements, which correspond to the TM($3d$) states 
and 
$F_{All}$ is the total sum squares of moduli elements.

%
%

It is clearly seen that the band at the Fermi level has this contribution dominant, 
whereas the rest valence band appears only moderately conditioned by TM($3d$) states.
We stress that as the qualitative evidence towards reported high sensitivity of the magnetic moment per TM atom to the lattice constants 
\cite{PhysRevB.80.014408, gopal2017improved}, 
taking into account known importance of electron-electron correlations in this context 
\cite{ZHANG20076}.
In our study for CoO (FeO) this moment is found as 
2.51~$\mu_B$ (3.58~$\mu_B$), 
which agrees well with the previous theoretical estimations 
\cite{PhysRevB.80.014408, cinthia2015first, gramsch2003structure}.
Nevertheless, we mention that these values appear slightly overestimated 
(in comparison with the experimental data 
\cite{PhysRev.110.1333, PhysRevB.65.125111}), due to essential dependence of TM $t_{2g}$ orbitals occupation from the spin-orbital coupling 
\cite{PhysRevB.86.115134}.

In case of both CoO and FeO we consider four TM atoms in the unit cell, which sets equal number of sublattices to be taken into account in MCA study.
If one freely chooses the TM atom to calculate MCA energy for, then the contribution from the sublattice containing this atom is assumed to capture intra-atomic physics along with inter-site effects driven by hopping integrals, whereas the rest three sublattices describe inter-site factors only.
In our case this atom is set as \#1, according to the local positions presented in~\ref{AppendixA}.
Thus we focus on the ferromagnetic (111) plane constituted by sublattices SL$_1$ and SL$_2$ (denoted as orange spheres in figure~\ref{fig:CrystalStructure}), and sublattices SL$_3$, SL$_4$ represent planes with alternating direction of the local magnetic moments
(gray spheres in figure~\ref{fig:CrystalStructure}).

The angular profiles of MCA energy are obtained using 
\EqRef{Eq:FinalMAE} 
on the grid 
$(\varphi \times \theta) = (8 \times 21)$.
It required about one thousand hours of computation on the modern multiprocessing computer. We thus emphasize that employment of $\bm{k}$-dependent Green's functions only provides technical possibility to perform such numerical experiments.
Whereas previously suggested approaches based on inter-site Green's functions 
\cite{PhysRevB.52.13419, PhysRevB.89.214422} are expected to take the time two orders of magnitude more 
\cite{kashin2024magnetocrystalline}, 
making the calculations nearly unimplementable by not only methodological, but also purely technical means.

The result of easy and hard axis determination in CoO and FeO is summarized in table~\ref{tab:MCA_Sublattices}.
In the study 
\cite{kashin2024magnetocrystalline} 
Orientation of the easy axis in (111) / along [111] was reported in CoO / FeO, and in present study we confirm that this plane plays exceptional role in MCA picture of both compounds. 
It becomes particularly evident as hard axes are obtained with nearly perfect direction in (111) plane and along [111] as well. 
Moreover, from CoO to FeO these axes are interchanged, which emphasizes the geometrical authenticity of this plane.

\begin{table}[htbp]
    \centering
    \caption
    {
    MCA energy of TM atom in CoO and FeO 
    (with local position \#1 in the unit cell, see~\ref{AppendixA}) 
    estimated using \EqRef{Eq:FinalMAE}, in meV. 
    Easy (hard) axis is determined by minimum (maximum) of the angular profile.
    Sublattices SL$_1$, SL$_2$ constitute (111) plane with the considered TM atom (orange spheres in figure~\ref{fig:CrystalStructure}), 
    and 
    SL$_3$, SL$_4$ represent (111) plane without this atom (gray spheres in figure~\ref{fig:CrystalStructure}).
    }
    \begin{tabular}{|c|c|c|c|c|c|c|}
        \hline
        \hline
         TMO & Direction \& Difference & Total value & SL$_1$ & SL$_2$ & SL$_3$ & SL$_4$ \\
        \hline
        \hline
               & (111) \{Easy\} & $-6.581$ & $-5.832$ & $-0.767$ & $0.002$ & $0.015$ \\
         \cline{2-7}
          CoO  & [111] \{Hard\} & $-3.901$ & $-3.895$ & $-0.009$ & $-0.009$ & $0.011$ \\
         \cline{2-7}
               & $\Delta$ = \{Hard\} $-$ \{Easy\} & $2.680$ & $1.937$ & $0.758$ & $-0.011$ & $-0.004$ \\
         \hline
               & (111) \{Hard\} & $-4.450$ & $-4.274$ & $-0.170$ & $-0.010$ & $0.004$ \\
         \cline{2-7}
         FeO   & [111] \{Easy\} & $-5.680$ & $-5.090$ & $-0.591$ & $-0.006$ & $0.007$ \\
         \cline{2-7}
               & $\Delta$ = \{Hard\} $-$ \{Easy\} & $1.231$ & $0.816$ & $0.421$ & $-0.004$ & $-0.003$ \\
         \hline
    \end{tabular}
    \label{tab:MCA_Sublattices}
\end{table}

As one can clearly see from MCA energy value description in terms of sublattices, the MCA picture in both CoO and FeO 
appears completely reasoned by only SL$_1$ and SL$_2$, 
which constitute ferromagnetic (111) plane containing the TM atom the angular profile is considered for, 
while SL$_3$ and SL$_4$ are found negligible.

It is also important to note that in case of hard axis both materials reveal smallness of SL$_2$ contribution, whereas the easy axis description and then MCA energy difference show this contribution commensurable with the SL$_1$.
Taking into account that the latter in general contains both intra-atomic and inter-site physics, we can draw a conclusion about hard axis development to be reasoned by intra-atomic effects dominantly, while kinetic effects driven by hopping integrals provide the energy lowering towards the easy axis.

\begin{figure}[htbp]
\begin{center}
\includegraphics[width=0.9\linewidth]{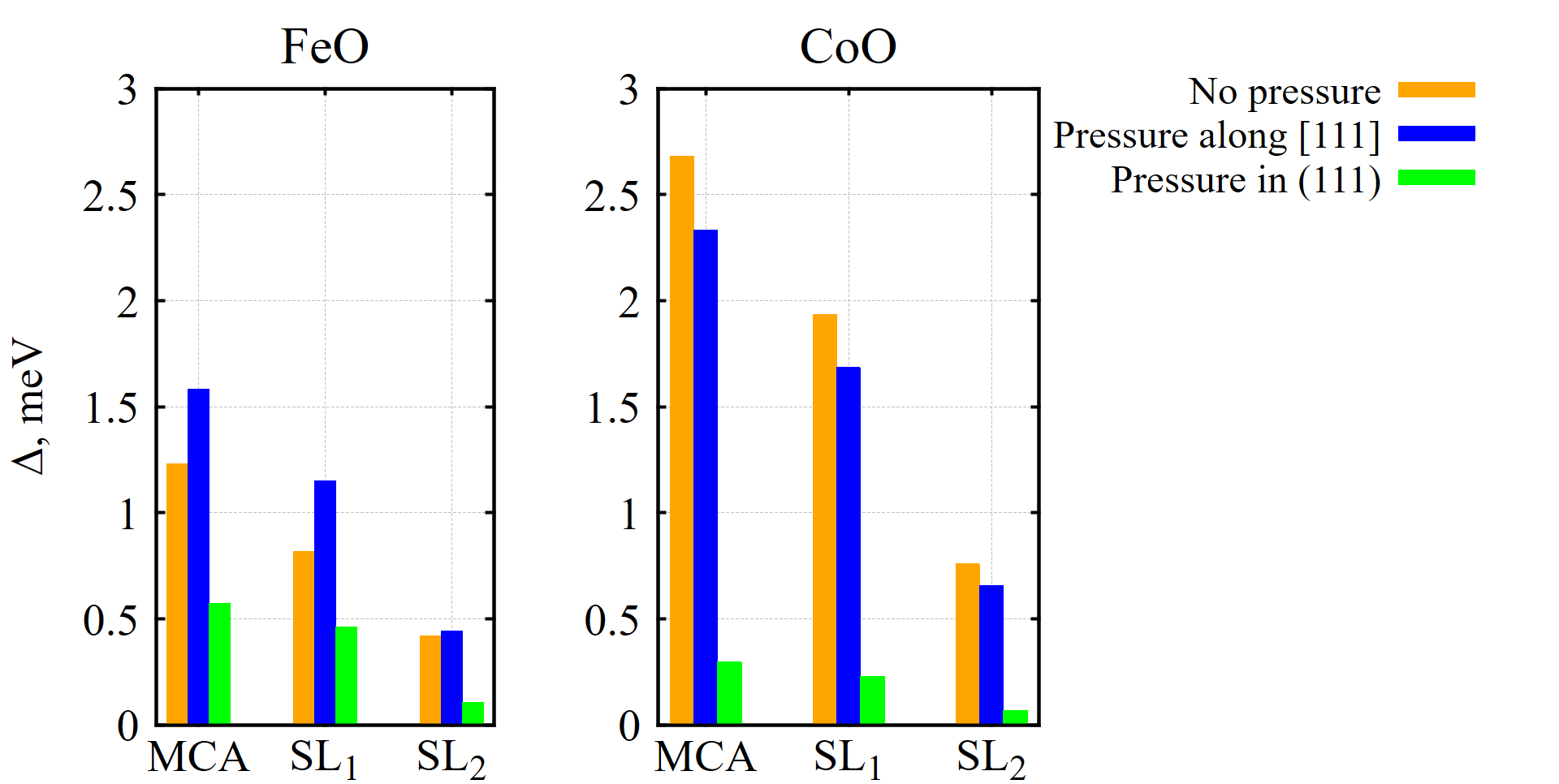} 
\caption
{
The MCA energy difference taken as 
$\Delta$ = \{Hard\} $-$ \{Easy\} 
 in CoO and FeO under the directional pressure in (111) and along [111], in meV. 
}
\label{fig:MCAPressure}
\end{center}
\end{figure}

This finding is in the established line of theoretical explanation based on the formation of long-range AFM II magnetic order, which is caused by TM$-$TM superexchange interaction and results in a crystal symmetry reduction 
\cite{PhysRevB.86.115134, PhysRev.79.350, ANDERSON196399, PhysRevB.80.014408}. 
In order to elaborate the question of the actual role that this interaction play we consider CoO and FeO under the directional pressure in (111) and along [111]. In terms of our model it means compression of the unit cell with decreasing of its size by 5\% in the corresponding direction, which could be expressed as the application of experimentally available $p ~\sim 10$ GPa 
\cite{PhysRevB.80.014408, Li2024}.
In figure~\ref{fig:MCAPressure} we present resulting MCA energy difference between hard and easy axis. 
One can clearly see that pressure in (111) significantly suppresses MCA, while the compression along [111] induces only slight deviation. 
Such contrast serves as additional and the most reliable evidence towards the role of (111) as being dominantly significant by means of the lattice geometry. 

%
%

\begin{figure}[htbp]
\begin{center}
\includegraphics[width=0.9\linewidth]{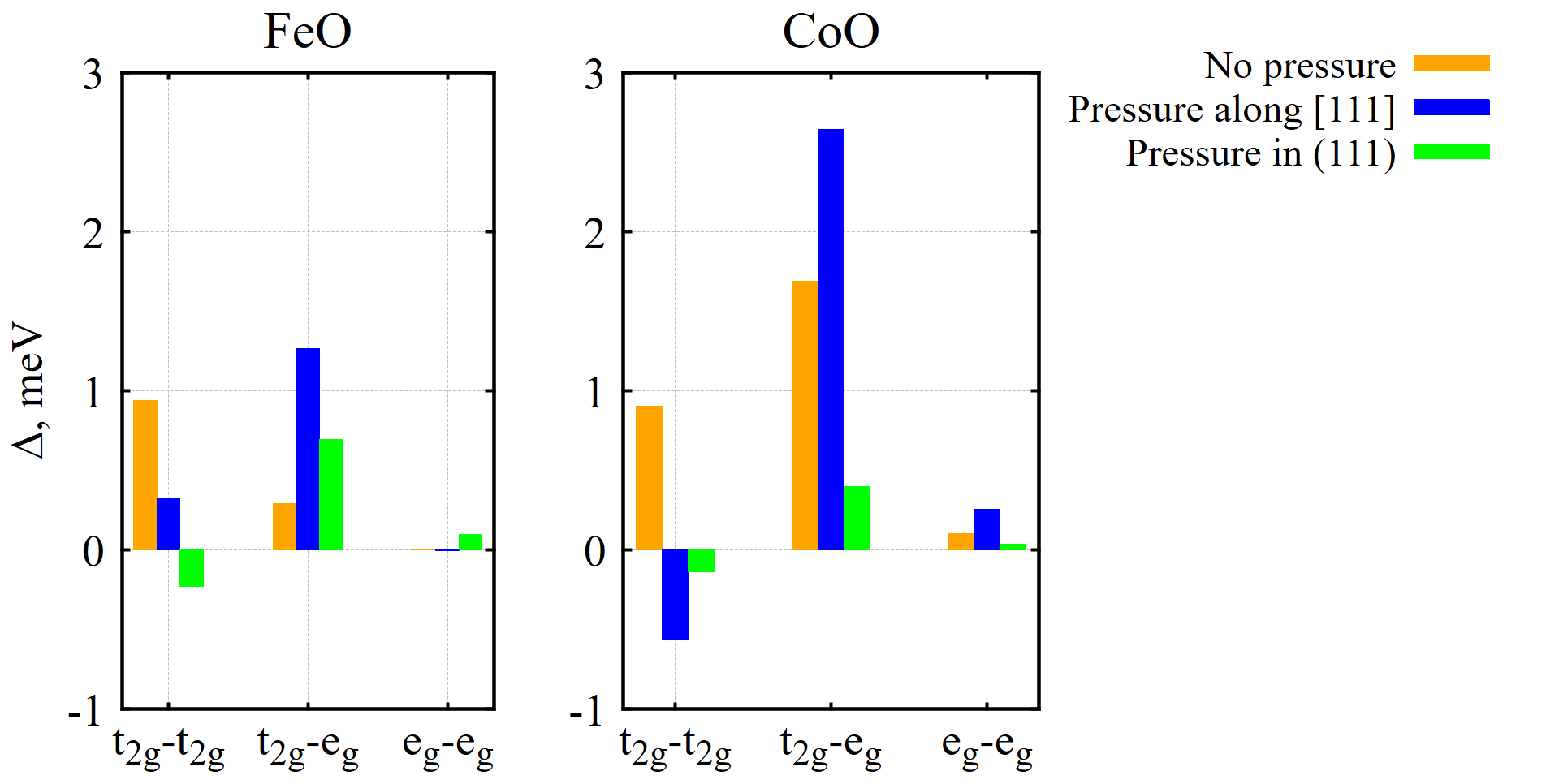} 
\caption
{
The orbital decomposition of MCA energy difference 
taken as $\Delta$~=~\{Hard\}~$-$~\{Easy\} 
in CoO and FeO under the directional pressure in (111) and along [111], in meV. 
}
\label{fig:MCAPressureOrbitalDecomposition}
\end{center}
\end{figure}

To trace the origins of this behavior we perform the orbital decomposition of MCA energy, \EqRef{Eq:MCADecomposition}, 
and of the exchange environment, 
\EqRef{Eq:JqDecomposed}, \EqRef{Eq:JiiDecomposed}, with the spin channels and the sublattices contributions summarized.
As one can see in figure~\ref{fig:MCAPressureOrbitalDecomposition} the dominant orbital contributions to MCA energy are reasoned by $t_{2g} - t_{2g}$ and $t_{2g} - e_{g}$ interactions, whereas the $e_{g} - e_{g}$ component is found negligible. 
Thus established decisive role of $t_{2g}$ orbitals confirms the Jahn-Teller effect to be the general driving force of MCA formation in CoO and FeO 
\cite{PhysRevB.86.115134, Jahn_Teller, Jahn_Bragg}. 

%
%

\begin{figure}[htbp]
\begin{center}
\includegraphics[width=0.9\linewidth]{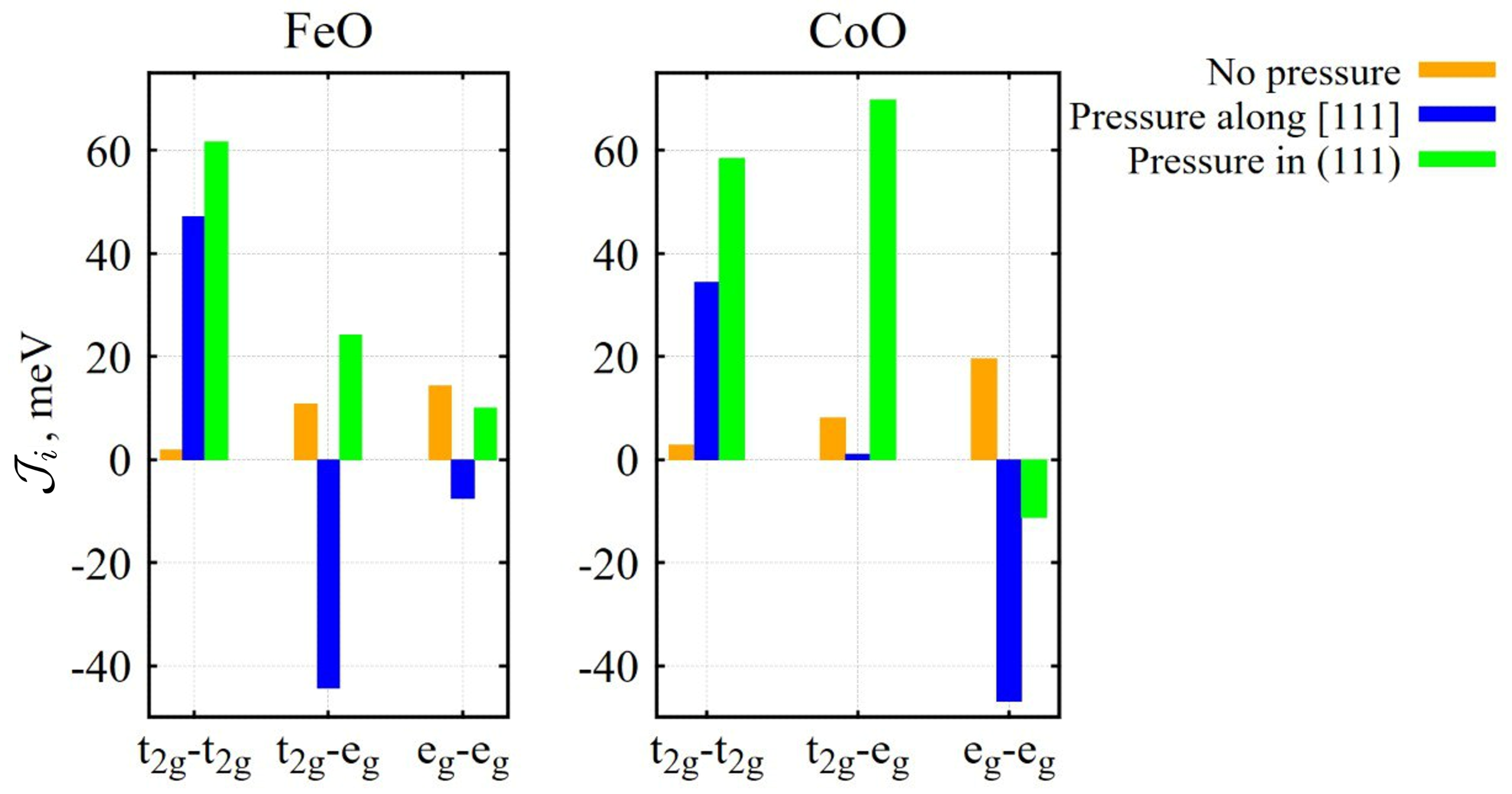} 
\caption
{
The orbital decomposition of the isotropic exchange environment, 
\EqRef{Eq:JqDecomposed}, \EqRef{Eq:JiiDecomposed}, 
taken as 
$ \{ {\cal{J}}_{i} \}^{\alpha \beta} = 
   \sum_{\sublattice{j}} \; \{ {\cal{J}}_{i, \; \sublattice{j}} \}^{\alpha \beta}$,
in CoO and FeO under the directional pressure in (111) and along [111], in meV. 
Positive (negative) values indicate the tendency towards AFM II (FM) order.
}
\label{fig:ExchangePressureOrbitalDecomposition}
\end{center}
\end{figure}

It appears remarkably fruitful to relate this decomposition with that found for the isotropic exchange environment in the form of 
$ \{ {\cal{J}}_{i} \}^{\alpha \beta} = 
   \sum_{\sublattice{j}} \; \{ {\cal{J}}_{i, \; \sublattice{j}} \}^{\alpha \beta}$ 
(figure~\ref{fig:ExchangePressureOrbitalDecomposition}).
We can clearly see that if the pressure induces extra tendencies towards stabilization of the AFM II order (positive values above orange columns), it leads to the suppression of the MCA.
Whereas if one observes the decay of these tendencies up to the switching of its character to FM (negative values), it results in essential boosting of the MCA energy difference.
Therefore we arrive to the conclusion that MCA effects are closely correlated with the isotropic exchange landscape, which reveals the magnetism in CoO and FeO, from one hand, as highly untypical since usually there is no straightforward way to provide a connection between these two magnetic phenomena in real magnetic material. 
From another hand, we thus reveal a strong sensitivity of the magnetic picture in CoO and FeO to mechanical influences, readily available in the experiments. 
It signifies new perspectives towards directional crystal field manipulation by means of not only pressure, but also synthesis of the thin films with robust anisotropic magnetism, applicable in the modern spintronics.

%
%





\section{Conclusion}

In the present study, we demonstrated that the role of the ferromagnetic plane (111) in the formation of the magnetocrystalline anisotropy picture in the monoxides FeO and CoO is determinative. The kinetic effects of the electrons in this plane dominantly control the orientation of the easy axis, and the robustness of this anisotropy can be stimulated or suppressed by directional pressure. We originally revealed the close interrelation between MCA and isotropic exchange environment, which opens up new prospects towards micromagnetic engineering driven by experimentally available tools of crystal field design.





\section{Acknowledgments}

The work is supported by the
Russian Science Foundation, 
Grant No. 23-72-01003,
https://rscf.ru/project/23-72-01003/







\appendix
\section{\textit{Ab initio} calculations of CoO and FeO}
\label{AppendixA}

The first-principles calculations of CoO and FeO electronic structure are carried out within density functional theory (DFT)~\cite{PhysRev.140.A1133}. 
In order to numerically realize it we use the generalized gradient approximation designed to take into account an effective Coulomb repulsion on the TM atoms (GGA$+U$).  
As the potential one commonly chooses the Perdew-Burke-Ernzerhof (PBE) exchange-correlation functional~\cite{PhysRevLett.77.3865}.
The Coulomb repulsion is based on the scheme of Dudarev~\textit{et~al.}~\cite{PhysRevB.57.1505}, which is implemented in the Quantum-Espresso simulation package~\cite{Giannozzi_2009} and employed in the present study.

The basic parameters of the calculations are following:
\begin{itemize}
\item The energy cutoff of the plane wave basis construction is 330~eV;
\item The energy convergence criterion is $10^{-6}$~eV;
\item The $20 \times 20 \times 20$ Monkhorst-Pack grid was employed to carry out integration over the 1st Brillouin zone;
\item The lattice vectors are:
$\bm{a}^{CoO}~=~(6.07, 0, 0)$~\AA; 
$\bm{b}^{CoO}~=~(0, 2.96, 0)$~\AA;
$\bm{c}^{CoO}~=~(3.08, 0, 4.23)$~\AA;
$\bm{a}^{FeO}~=~(6.19, 0, 0)$~\AA; 
$\bm{b}^{FeO}~=~(0, 3.02, 0)$~\AA;
$\bm{c}^{FeO}~=~(3.11, 0, 4.48)$~\AA;
\item The local positions of four TM atoms in the unit cell are:
$\bm{r}^{CoO}_1~=~(0, 0, 0)$~\AA;
$\bm{r}^{CoO}_2~=~(4.57, 1.48, 2.11)$~\AA;
$\bm{r}^{CoO}_3~=~(3.03, 0, 0)$~\AA;
$\bm{r}^{CoO}_4~=~(1.54, 1.48, 2.11)$~\AA;
$\bm{r}^{FeO}_1~=~(0, 0, 0)$~\AA;
$\bm{r}^{FeO}_2~=~(4.65, 1.51, 2.24)$~\AA;
$\bm{r}^{FeO}_3~=~(3.09, 0, 0)$~\AA;
$\bm{r}^{FeO}_4~=~(1.56, 1.51, 2.24)$~\AA;
\item An effective Coulomb repulsion parameter $U~=~4$~eV, in a full agreement with~\cite{PhysRevB.86.115134}.
\end{itemize}

The resulting band structure is presented in figure~\ref{fig:BandStructure} and is consistent with the previous studies 
\cite{PhysRevB.49.10170, PhysRevB.74.155108, i9061050}. 
The bands near Fermi level are due to hybridization of TM $3d$ states and oxygen $2p$ states, as well as by the strong localization of the electrons on TM $3d$ states~\cite{PhysRevB.86.115134, cinthia2015first}.

In order to build the low-energy model of magnetoactive TM $3d$-shell we perform a projection of the DFT wave functions onto maximally localized Wannier functions
~\cite{PhysRevB.56.12847, PhysRevB.65.035109, MOSTOFI2008685}, 
which capture along with the $3d$ states also $4s,~4p$ states for TM and $2s,~2p$ states for oxygen, due to the essential entanglement of corresponding bands. 
The resulting model perfectly reproduces DFT band structure in wide energy range.

The small parameter in the SOC operator $\lambda =$ 0.02~eV, being adjusted for a projection factor due to $^{4}T_{1}$ ground state at low temperatures 
\cite{PhysRevB.88.205117, Sarte_2020, SANCHEZNAVAS20042261}.






\section*{References}

\bibliographystyle{iopart-num-long}

\bibliography{Biblio}

\providecommand{\newblock}{}
\begin{thebibliography}{10}
\expandafter\ifx\csname url\endcsname\relax
  \def\url#1{{\tt #1}}\fi
\expandafter\ifx\csname urlprefix\endcsname\relax\def\urlprefix{URL }\fi
\providecommand{\eprint}[2][arXiv]{#1:\linebreak[0]#2}

\bibitem{ohara2021confinement}
Ohara K, Zhang X, Chen Y, Wei Z, Ma Y, Xia J, Zhou Y and Liu X 2021 Confinement and protection of skyrmions by patterns of modified magnetic properties {\em Nano Lett.\/} {\bf 21} 4320--4326

\bibitem{ohara2022reversible}
Ohara K, Zhang X, Chen Y, Kato S, Xia J, Ezawa M, Tretiakov O~A, Hou Z, Zhou Y, Zhao G {\em et~al\/} 2022 Reversible transformation between isolated skyrmions and bimerons {\em Nano Lett.\/} {\bf 22} 8559--8566

\bibitem{hassan2024dipolar}
Hassan M, Koraltan S, Ullrich A, Bruckner F, Serha R~O, Levchenko K~V, Varvaro G, Kiselev N~S, Heigl M, Abert C {\em et~al\/} 2024 Dipolar skyrmions and antiskyrmions of arbitrary topological charge at room temperature {\em Nat. Phys.\/}  1--8

\bibitem{PhysRevMaterials.2.051004}
Lin Z, Lohmann M, Ali Z~A, Tang C, Li J, Xing W, Zhong J, Jia S, Han W, Coh S, Beyermann W and Shi J 2018 Pressure-induced spin reorientation transition in layered ferromagnetic insulator {Cr}$_2${Ge}$_2${Te}$_6$ {\em Phys. Rev. Mater.\/} {\bf 2}(5) 051004 \urlprefix\url{https://link.aps.org/doi/10.1103/PhysRevMaterials.2.051004}

\bibitem{D1DT01719E}
Briganti M and Totti F 2021 Magnetic anisotropy on demand exploiting high-pressure as remote control: an ab initio proof of concept {\em Dalton Trans.\/} {\bf 50}(30) 10621--10628 \urlprefix\url{http://dx.doi.org/10.1039/D1DT01719E}

\bibitem{craig2018probing}
Craig G~A, Sarkar A, Woodall C~H, Hay M~A, Marriott K~E, Kamenev K~V, Moggach S~A, Brechin E~K, Parsons S, Rajaraman G {\em et~al\/} 2018 Probing the origin of the giant magnetic anisotropy in trigonal bipyramidal {Ni (II)} under high pressure {\em Chem. Sci.\/} {\bf 9} 1551--1559

\bibitem{meng2016understanding}
Meng Y~S, Jiang S~D, Wang B~W and Gao S 2016 Understanding the magnetic anisotropy toward single-ion magnets {\em Acc. Chem. Res.\/} {\bf 49} 2381--2389

\bibitem{yuan2013spin}
Yuan H, Chen H, Kuang A and Wu B 2013 Spin--orbit effect and magnetic anisotropy in {Pt} clusters {\em J. Magn. Magn. Mater.\/} {\bf 331} 7--16

\bibitem{PhysRevB.79.224418}
B\l{}o\ifmmode~\acute{n}\else \'{n}\fi{}ski P and Hafner J 2009 Magnetic anisotropy of transition-metal dimers: {Density} functional calculations {\em Phys. Rev. B\/} {\bf 79}(22) 224418 \urlprefix\url{https://link.aps.org/doi/10.1103/PhysRevB.79.224418}

\bibitem{blonski2011magneto}
B{\l}o{\'n}ski P and Hafner J 2011 Magneto-structural properties and magnetic anisotropy of small transition-metal clusters: a first-principles study {\em J. Phys. Condens. Matter\/} {\bf 23} 136001

\bibitem{bar2016magnetic}
Bar A~K, Pichon C and Sutter J~P 2016 Magnetic anisotropy in two-to eight-coordinated transition--metal complexes: Recent developments in molecular magnetism {\em Coord. Chem. Rev.\/} {\bf 308} 346--380

\bibitem{C4CC05724D}
Saber M~R and Dunbar K~R 2014 Ligands effects on the magnetic anisotropy of tetrahedral cobalt complexes {\em Chem. Commun.\/} {\bf 50}(82) 12266--12269 \urlprefix\url{http://dx.doi.org/10.1039/C4CC05724D}

\bibitem{parker2020ligand}
Parker D, Suturina E~A, Kuprov I and Chilton N~F 2020 How the ligand field in lanthanide coordination complexes determines magnetic susceptibility anisotropy, paramagnetic nmr shift, and relaxation behavior {\em Acc. Chem. Res.\/} {\bf 53} 1520--1534

\bibitem{goswami2012ligand}
Goswami T and Misra A 2012 Ligand effects toward the modulation of magnetic anisotropy and design of magnetic systems with desired anisotropy characteristics {\em J. Phys. Chem. A\/} {\bf 116} 5207--5215

\bibitem{schweinfurth2017tuning}
Schweinfurth D, Krzystek J, Atanasov M, Klein J, Hohloch S, Telser J, Demeshko S, Meyer F, Neese F and Sarkar B 2017 Tuning magnetic anisotropy through ligand substitution in five-coordinate {Co (II)} complexes {\em Inorg. Chem.\/} {\bf 56} 5253--5265

\bibitem{lado2017origin}
Lado J~L and Fern{\'a}ndez-Rossier J 2017 On the origin of magnetic anisotropy in two dimensional {CrI}$_3$ {\em 2D Mater.\/} {\bf 4} 035002

\bibitem{TwoDimensionalSystems_Qu}
Qu Y, Liao Y, He J, Chen Y and Yao G 2024 High-temperature intrinsic two-dimensional-{XY} ferromagnetism and strong magnetoelastic coupling in tetragonal monolayer {MnGe} {\em J. Phys. Chem. C\/} {\bf 128} 4631--4638 \urlprefix\url{https://doi.org/10.1021/acs.jpcc.3c06990}

\bibitem{PhysRevMaterials.6.L061002}
Dey D, Ray A and Yu L 2022 Intrinsic ferromagnetism and restrictive thermodynamic stability in {MA}$_2${N}$_4$ and {Janus} {VSiGeN}$_4$ monolayers {\em Phys. Rev. Mater.\/} {\bf 6}(6) L061002 \urlprefix\url{https://link.aps.org/doi/10.1103/PhysRevMaterials.6.L061002}

\bibitem{PhysRevApplied.3.034009}
Hao Q and Xiao G 2015 Giant spin {Hall} effect and switching induced by spin-transfer torque in a {W}/{Co}$_{40}${Fe}$_{40}${B}$_{20}$/{MgO} structure with perpendicular magnetic anisotropy {\em Phys. Rev. Appl.\/} {\bf 3}(3) 034009 \urlprefix\url{https://link.aps.org/doi/10.1103/PhysRevApplied.3.034009}

\bibitem{almasi2015enhanced}
Almasi H, Hickey D~R, Newhouse-Illige T, Xu M, Rosales M, Nahar S, Held J, Mkhoyan K and Wang W 2015 Enhanced tunneling magnetoresistance and perpendicular magnetic anisotropy in {Mo/CoFeB/MgO} magnetic tunnel junctions {\em Appl. Phys. Lett.\/} {\bf 106}

\bibitem{stohr1999exploring}
St{\"o}hr J 1999 Exploring the microscopic origin of magnetic anisotropies with {X-ray} magnetic circular dichroism {(XMCD)} spectroscopy {\em J. Magn. Magn. Mater.\/} {\bf 200} 470--497

\bibitem{shepley2015modification}
Shepley P, Rushforth A, Wang M, Burnell G and Moore T 2015 Modification of perpendicular magnetic anisotropy and domain wall velocity in {Pt/Co/Pt} by voltage-induced strain {\em Sci. Rep.\/} {\bf 5} 7921

\bibitem{bilz1979metal}
Bilz H, Kress W, Bilz H and Kress W 1979 Metal oxides (rock salt structure) {\em Phonon Dispersion Relations in Insulators\/}  49--57

\bibitem{10.1063/5.0082729}
Kozioł-Rachwał A, Szpytma M, Spiridis N, Freindl K, Korecki J, Janus W, Nayyef H, Dróżdż P, Ślęzak M, Zając M and Ślęzak T 2022 Beating the limitation of the néel temperature of {FeO} with antiferromagnetic proximity in {FeO/CoO} {\em Appl. Phys. Lett.\/} {\bf 120} 072404 ISSN 0003-6951 \urlprefix\url{https://doi.org/10.1063/5.0082729}

\bibitem{PhysRevB.86.115134}
Schr\"on A, R\"odl C and Bechstedt F 2012 Crystalline and magnetic anisotropy of the 3$d$-transition metal monoxides {MnO, FeO, CoO, and NiO} {\em Phys. Rev. B\/} {\bf 86}(11) 115134 \urlprefix\url{https://link.aps.org/doi/10.1103/PhysRevB.86.115134}

\bibitem{LIU201796}
Liu J~J, Meng Y, Ren P, Zhaorigetu B, Guo W, Cao D~B, Li Y~W, Jiao H, Liu Z, Jia M, Yang Y, Xu A and Wen X~D 2017 Predicting the structural and electronic properties of transition metal monoxides from bulk to surface morphology {\em Catal. Today\/} {\bf 282} 96--104 ISSN 0920-5861 \urlprefix\url{https://www.sciencedirect.com/science/article/pii/S0920586116305090}

\bibitem{parida2018universality}
Parida P, Kashikar R, Jena A and Nanda B 2018 Universality in the electronic structure of 3$d$ transition metal oxides {\em J. Phys. Chem. Solids\/} {\bf 123} 133--149

\bibitem{deng2010origin}
Deng H~X, Li J, Li S~S, Xia J~B, Walsh A and Wei S~H 2010 Origin of antiferromagnetism in {CoO}: {A} density functional theory study {\em Appl. Phys. Lett.\/} {\bf 96}

\bibitem{PhysRev.79.350}
Anderson P~W 1950 Antiferromagnetism. {Theory} of superexchange interaction {\em Phys. Rev.\/} {\bf 79}(2) 350--356 \urlprefix\url{https://link.aps.org/doi/10.1103/PhysRev.79.350}

\bibitem{ANDERSON196399}
Anderson P~W 1963 {\em Theory of Magnetic Exchange Interactions: {Exchange} in Insulators and Semiconductors\/} ({\em Solid State Physics\/} vol~14) (Academic Press) pp 99--214 \urlprefix\url{https://www.sciencedirect.com/science/article/pii/S008119470860260X}

\bibitem{PhysRevB.80.014408}
Fischer G, D\"ane M, Ernst A, Bruno P, L\"uders M, Szotek Z, Temmerman W and Hergert W 2009 Exchange coupling in transition metal monoxides: {Electronic} structure calculations {\em Phys. Rev. B\/} {\bf 80}(1) 014408 \urlprefix\url{https://link.aps.org/doi/10.1103/PhysRevB.80.014408}

\bibitem{PhysRev.110.1333}
Roth W~L 1958 Magnetic structures of {MnO, FeO, CoO, and NiO} {\em Phys. Rev.\/} {\bf 110}(6) 1333--1341 \urlprefix\url{https://link.aps.org/doi/10.1103/PhysRev.110.1333}

\bibitem{herrmann1978equivalent}
Herrmann-Ronzaud D, Burlet P and Rossat-Mignod J 1978 Equivalent {type-II} magnetic structures: {CoO}, a collinear antiferromagnet {\em J. Phys. C: Solid State Phys.\/} {\bf 11} 2123

\bibitem{PhysRevB.64.052102}
Jauch W, Reehuis M, Bleif H~J, Kubanek F and Pattison P 2001 Crystallographic symmetry and magnetic structure of {CoO} {\em Phys. Rev. B\/} {\bf 64}(5) 052102 \urlprefix\url{https://link.aps.org/doi/10.1103/PhysRevB.64.052102}

\bibitem{fjellvag2002monoclinic}
Fjellv{\aa}g H, Hauback B~C, Vogt T and St{\o}len S 2002 Monoclinic nearly stoichiometric wustite at low temperatures {\em Am. Min.\/} {\bf 87} 347--349

\bibitem{Jahn_Teller}
Jahn H~A, Teller E and Donnan F~G 1937 Stability of polyatomic molecules in degenerate electronic states - {I-Orbital} degeneracy {\em Proc. R. Soc. London Ser. A\/} {\bf 161} 220--235 \urlprefix\url{https://royalsocietypublishing.org/doi/abs/10.1098/rspa.1937.0142}

\bibitem{Jahn_Bragg}
Jahn H~A and Bragg W~H 1938 Stability of polyatomic molecules in degenerate electronic states {II-Spin} degeneracy {\em Proc. R. Soc. London Ser. A\/} {\bf 164} 117--131 \urlprefix\url{https://royalsocietypublishing.org/doi/abs/10.1098/rspa.1938.0008}

\bibitem{schron2013magnetic}
Schr{\"o}n A and Bechstedt F 2013 Magnetic anisotropy of {FeO} and {CoO}: the influence of gradient corrections on exchange and correlation {\em J. Phys. Condens. Matter\/} {\bf 25} 486002

\bibitem{gopal2017improved}
Gopal P, De~Gennaro R, dos Santos~Gusmao M~S, Al~Orabi R~A~R, Wang H, Curtarolo S, Fornari M and Nardelli M~B 2017 Improved electronic structure and magnetic exchange interactions in transition metal oxides {\em J. Phys. Condens. Matter\/} {\bf 29} 444003

\bibitem{tomiyasu2006magnetic}
Tomiyasu K and Itoh S 2006 Magnetic excitations in {CoO} {\em JPSJ\/} {\bf 75} 084708

\bibitem{PhysRevB.18.1317}
Kugel G~E, Hennion B and Carabatos C 1978 Low-energy magnetic excitations in wustite ({Fe}$_{1-x}${O}) {\em Phys. Rev. B\/} {\bf 18}(3) 1317--1321 \urlprefix\url{https://link.aps.org/doi/10.1103/PhysRevB.18.1317}

\bibitem{rechtin1972long}
Rechtin M and Averbach B 1972 Long-range magnetic order in {CoO} {\em Phys. Rev. B\/} {\bf 6} 4294

\bibitem{PhysRevLett.77.3865}
Perdew J~P, Burke K and Ernzerhof M 1996 Generalized gradient approximation made simple {\em Phys. Rev. Lett.\/} {\bf 77}(18) 3865--3868 \urlprefix\url{https://link.aps.org/doi/10.1103/PhysRevLett.77.3865}

\bibitem{PhysRevB.52.13419}
Solovyev I~V, Dederichs P~H and Mertig I 1995 Origin of orbital magnetization and magnetocrystalline anisotropy in {TX} ordered alloys (where {T=Fe,Co} and {X=Pd,Pt}) {\em Phys. Rev. B\/} {\bf 52}(18) 13419--13428 \urlprefix\url{https://link.aps.org/doi/10.1103/PhysRevB.52.13419}

\bibitem{PhysRevB.89.214422}
Mazurenko V~V, Kvashnin Y~O, Jin F, De~Raedt H~A, Lichtenstein A~I and Katsnelson M~I 2014 First-principles modeling of magnetic excitations in {Mn}$_{12}$ {\em Phys. Rev. B\/} {\bf 89}(21) 214422 \urlprefix\url{https://link.aps.org/doi/10.1103/PhysRevB.89.214422}

\bibitem{cinthia2015first}
Cinthia A~J, Rajeswarapalanichamy R and Iyakutti K 2015 First principles study of electronic structure, magnetic, and mechanical properties of transition metal monoxides {TMO(TM=Co and Ni)} {\em Zeitschrift für Naturforschung A\/} {\bf 70} 797--804 \urlprefix\url{https://doi.org/10.1515/zna-2015-0216}

\bibitem{kashin2024magnetocrystalline}
Kashin I~V and Andreev S~N 2024 Magnetocrystalline anisotropy in metallic systems: fast and stable estimation in greens functions formalism {\em arXiv preprint arXiv:2403.14241\/}

\bibitem{LIECHTENSTEIN198765}
Liechtenstein A, Katsnelson M, Antropov V and Gubanov V 1987 Local spin density functional approach to the theory of exchange interactions in ferromagnetic metals and alloys {\em J. Magn. Magn. Mater.\/} {\bf 67} 65--74 ISSN 0304-8853 \urlprefix\url{https://www.sciencedirect.com/science/article/pii/0304885387907219}

\bibitem{PhysRevB.106.134434}
Kashin I~V, Gerasimov A and Mazurenko V~V 2022 Reciprocal space study of {Heisenberg} exchange interactions in ferromagnetic metals {\em Phys. Rev. B\/} {\bf 106}(13) 134434 \urlprefix\url{https://link.aps.org/doi/10.1103/PhysRevB.106.134434}

\bibitem{PhysRevB.13.5188}
Monkhorst H~J and Pack J~D 1976 Special points for {Brillouin-zone} integrations {\em Phys. Rev. B\/} {\bf 13}(12) 5188--5192 \urlprefix\url{https://link.aps.org/doi/10.1103/PhysRevB.13.5188}

\bibitem{PhysRevB.39.865}
Bruno P 1989 Tight-binding approach to the orbital magnetic moment and magnetocrystalline anisotropy of transition-metal monolayers {\em Phys. Rev. B\/} {\bf 39}(1) 865--868 \urlprefix\url{https://link.aps.org/doi/10.1103/PhysRevB.39.865}

\bibitem{PhysRevB.47.14932}
Wang D~s, Wu R and Freeman A~J 1993 First-principles theory of surface magnetocrystalline anisotropy and the diatomic-pair model {\em Phys. Rev. B\/} {\bf 47}(22) 14932--14947 \urlprefix\url{https://link.aps.org/doi/10.1103/PhysRevB.47.14932}

\bibitem{Goringe_1997}
Goringe C~M, Bowler D~R and Hernández E 1997 Tight-binding modelling of materials {\em Rep. Prog. Phys.\/} {\bf 60} 1447 \urlprefix\url{https://dx.doi.org/10.1088/0034-4885/60/12/001}

\bibitem{PhysRevB.91.125133}
Kvashnin Y~O, Gr\aa{}n\"as O, Di~Marco I, Katsnelson M~I, Lichtenstein A~I and Eriksson O 2015 Exchange parameters of strongly correlated materials: {Extraction} from spin-polarized density functional theory plus dynamical mean-field theory {\em Phys. Rev. B\/} {\bf 91}(12) 125133 \urlprefix\url{https://link.aps.org/doi/10.1103/PhysRevB.91.125133}

\bibitem{LocalForceTheorem_1}
Machintosh A and Andersen O 1980 {\em Electrons at the {Fermi} surface\/} (Cambridge University Press) p 149

\bibitem{LocalForceTheorem_2}
Methfessel M and Kubler J 1982 Bond analysis of heats of formation: application to some group {VIII} and {IB} hydrides {\em J. Phys. F: Met. Phys.\/} {\bf 12} 141--161 \urlprefix\url{https://doi.org/10.1088/0305-4608/12/1/013}

\bibitem{Lichtenstein2013correl13}
Lichtenstein A 2013 {\em Magnetism: {From} {S}toner to {H}ubbard\/} ({Forschungszentrum Jülich GmbH Institute for Advanced Simulations, Jülich, Germany}) \urlprefix\url{https://www.cond-mat.de/events/correl13/manuscripts/lichtenstein.pdf}

\bibitem{PhysRevLett.116.217202}
Kvashnin Y~O, Cardias R, Szilva A, Di~Marco I, Katsnelson M~I, Lichtenstein A~I, Nordstr\"om L, Klautau A~B and Eriksson O 2016 Microscopic origin of {Heisenberg} and {Non-Heisenberg} exchange interactions in ferromagnetic bcc {Fe} {\em Phys. Rev. Lett.\/} {\bf 116}(21) 217202 \urlprefix\url{https://link.aps.org/doi/10.1103/PhysRevLett.116.217202}

\bibitem{PhysRevB.30.4734}
Terakura K, Oguchi T, Williams A~R and K\"ubler J 1984 Band theory of insulating transition-metal monoxides: {Band-structure} calculations {\em Phys. Rev. B\/} {\bf 30}(8) 4734--4747 \urlprefix\url{https://link.aps.org/doi/10.1103/PhysRevB.30.4734}

\bibitem{PhysRevB.49.10170}
Dufek P, Blaha P, Sliwko V and Schwarz K 1994 Generalized-gradient-approximation description of band splittings in transition-metal oxides and fluorides {\em Phys. Rev. B\/} {\bf 49}(15) 10170--10175 \urlprefix\url{https://link.aps.org/doi/10.1103/PhysRevB.49.10170}

\bibitem{PhysRevB.74.155108}
Tran F, Blaha P, Schwarz K and Nov\'ak P 2006 Hybrid exchange-correlation energy functionals for strongly correlated electrons: {Applications} to transition-metal monoxides {\em Phys. Rev. B\/} {\bf 74}(15) 155108 \urlprefix\url{https://link.aps.org/doi/10.1103/PhysRevB.74.155108}

\bibitem{ZHANG20076}
Zhang W~B, Deng Y~H, Hu Y~L, Han K~L and Tang B~Y 2007 Structural distortion of b1-structured {MnO} and {FeO} {\em Solid State Commun.\/} {\bf 142} 6--9 ISSN 0038-1098 \urlprefix\url{https://www.sciencedirect.com/science/article/pii/S0038109807000920}

\bibitem{gramsch2003structure}
Gramsch S~A, Cohen R~E and Savrasov S~Y 2003 Structure, metal-insulator transitions, and magnetic properties of {FeO} at high pressures {\em Am. Min.\/} {\bf 88} 257--261 \urlprefix\url{https://doi.org/10.2138/am-2003-2-301}

\bibitem{PhysRevB.65.125111}
Jauch W and Reehuis M 2002 Electron density distribution in paramagnetic and antiferromagnetic {CoO}: {A} $\gamma$-ray diffraction study {\em Phys. Rev. B\/} {\bf 65}(12) 125111 \urlprefix\url{https://link.aps.org/doi/10.1103/PhysRevB.65.125111}

\bibitem{Li2024}
Li X, Bykova E, Vasiukov D, Aprilis G, Chariton S, Cerantola V, Bykov M, M{\"u}ller S, Pakhomova A, Akbar F~I, Mukhina E, Kantor I, Glazyrin K, Comboni D, Chumakov A~I, McCammon C, Dubrovinsky L, Sanchez-Valle C and Kupenko I 2024 Monoclinic distortion and magnetic transitions in feo under pressure and temperature {\em Communications Physics\/} {\bf 7} 305 ISSN 2399-3650 \urlprefix\url{https://doi.org/10.1038/s42005-024-01797-1}

\bibitem{PhysRev.140.A1133}
Kohn W and Sham L~J 1965 Self-consistent equations including exchange and correlation effects {\em Phys. Rev.\/} {\bf 140}(4A) A1133--A1138 \urlprefix\url{https://link.aps.org/doi/10.1103/PhysRev.140.A1133}

\bibitem{PhysRevB.57.1505}
Dudarev S~L, Botton G~A, Savrasov S~Y, Humphreys C~J and Sutton A~P 1998 {Electron-energy-loss spectra and the structural stability of nickel oxide: An LSDA+U study} {\em Phys. Rev. B\/} {\bf 57}(3) 1505--1509 \urlprefix\url{https://link.aps.org/doi/10.1103/PhysRevB.57.1505}

\bibitem{Giannozzi_2009}
Giannozzi P, Baroni S, Bonini N, Calandra M, Car R, Cavazzoni C, Ceresoli D, Chiarotti G~L, Cococcioni M, Dabo I, Corso A~D, de~Gironcoli S, Fabris S, Fratesi G, Gebauer R, Gerstmann U, Gougoussis C, Kokalj A, Lazzeri M, Martin-Samos L, Marzari N, Mauri F, Mazzarello R, Paolini S, Pasquarello A, Paulatto L, Sbraccia C, Scandolo S, Sclauzero G, Seitsonen A~P, Smogunov A, Umari P and Wentzcovitch R~M 2009 {QUANTUM ESPRESSO: a modular and open-source software project for quantum simulations of materials} {\em J. Phys.: Condens. Matter\/} {\bf 21} 395502 \urlprefix\url{https://dx.doi.org/10.1088/0953-8984/21/39/395502}

\bibitem{i9061050}
Putz M~V 2008 Density functionals of chemical bonding {\em Int. J. Mol. Sci.\/} {\bf 9} 1050--1095 ISSN 1422-0067 \urlprefix\url{https://www.mdpi.com/1422-0067/9/6/1050}

\bibitem{PhysRevB.56.12847}
Marzari N and Vanderbilt D 1997 {Maximally localized generalized Wannier functions for composite energy bands} {\em Phys. Rev. B\/} {\bf 56}(20) 12847--12865 \urlprefix\url{https://link.aps.org/doi/10.1103/PhysRevB.56.12847}

\bibitem{PhysRevB.65.035109}
Souza I, Marzari N and Vanderbilt D 2001 {Maximally localized Wannier functions for entangled energy bands} {\em Phys. Rev. B\/} {\bf 65}(3) 035109 \urlprefix\url{https://link.aps.org/doi/10.1103/PhysRevB.65.035109}

\bibitem{MOSTOFI2008685}
Mostofi A~A, Yates J~R, Lee Y~S, Souza I, Vanderbilt D and Marzari N 2008 {wannier90: A tool for obtaining maximally-localised Wannier functions} {\em Comput. Phys. Commun.\/} {\bf 178} 685--699 ISSN 0010-4655 \urlprefix\url{https://www.sciencedirect.com/science/article/pii/S0010465507004936}

\bibitem{PhysRevB.88.205117}
Cowley R~A, Buyers W~J~L, Stock C, Yamani Z, Frost C, Taylor J~W and Prabhakaran D 2013 Neutron scattering investigation of the $d-d$ excitations below the {Mott} gap of {CoO} {\em Phys. Rev. B\/} {\bf 88}(20) 205117 \urlprefix\url{https://link.aps.org/doi/10.1103/PhysRevB.88.205117}

\bibitem{Sarte_2020}
Sarte P~M, Wilson S~D, Attfield J~P and Stock C 2020 Magnetic fluctuations and the spin–orbit interaction in {Mott} insulating {CoO} {\em J. Phys.: Condens. Matter\/} {\bf 32} 374011 \urlprefix\url{https://dx.doi.org/10.1088/1361-648X/ab8498}

\bibitem{SANCHEZNAVAS20042261}
{Sanchez Navas} A, Reddy B and Nieto F 2004 Spectroscopic study of chromium, iron, {OH}, fluid and mineral inclusions in uvarovite and fuchsite {\em Spectrochim. Acta A Mol. Biomol. Spectrosc.\/} {\bf 60} 2261--2268 ISSN 1386-1425 \urlprefix\url{https://www.sciencedirect.com/science/article/pii/S1386142503005523}

\end{thebibliography}



\end{document}